\documentclass[aps,pra,superscriptaddress]{revtex4}
\usepackage{graphicx}
\usepackage{amssymb}
\usepackage{amsmath}
\usepackage[retainorgcmds]{IEEEtrantools}
\usepackage{amsfonts}

\begin{document}
\Large

\preprint{}
\title{ Simulation of a Partially Entangled Two Qubit State Correlation with one PR-Box and one M-box}
\author{Ali Ahanj$^1$}
\email{a.ahanj@khayyam.ac.ir ; ahanj@ipm.ir}
\author{Pramod S. Joag$^2$}
\email{pramod@physics.unipune.ac.in}
\affiliation{$^{1}$Department of Physics, Khayyam Institute of Higher Education,  Mashhad, Iran and\\ School of Physics, Institute for Research in Fundamental Science(IPM), P. O. Box 19395-5531, Tehran, Iran.\\$^{2}$ Department of Physics, University of Pune, Pune-411007, India.}
\begin{abstract}
We present a protocol to simulate the quantum correlation implied by non
maximally entangled two qubit states, in the worst case scenario. This protocol
makes a single use of PR-box and a single use of Millionaire box (M-box). To the
best of our knowledge, the resources used in this protocol are weaker than those used in
previous protocols and are minimal  in the worst case scenario.
\end{abstract}
\maketitle
 \section{Introduction}
 One of the most intriguing features of quantum physics is the
 non-locality of correlations obtained by measuring entangled quantum
 particles. These correlations are nonlocal because they are neither
 caused by an exchange of a signal, as any hypothetical signal should
 travel faster than light, nor are they due to any pre-determined
 agreement (shared randomness) as they break Bell's inequalities
 \cite{bell}. A natural way to understand these correlations is to
 classically simulate them ( known as simulation of entanglement)
 using minimal resources. Obviously, this cannot be done using only
 local resources, that is, using shared randomness.
 The local resources must be supplemented by non-local ones. A simple non-local
 resource is communication of information via classical bits (cbits),
 which we can quantify and thus provides a `measure of non-locality'.
 In this scenario, Alice and Bob try and output $\alpha$ and $\beta$
 respectively, through a classical protocol, with the same
 probability distribution as if they shared the bipartite entangled
 quantum system and each measured his or her part of the system according to
 a given random Von Neumann measurement. As we have mentioned above,
 such a protocol must involve communication between Alice and Bob,
 who generally share finite or infinite number of random variables.
 The amount of communication is quantified \cite{pironio} either as
 the average number of cbits C(P) over the directions along which the
 spin components are measured (average or expected communication) or
 the worst case communication, which is  the maximum amount of
 communication $C_w(P)$ exchanged between Alice and Bob in any
 particular execution of the protocol. The third method is asymptotic
 communication i.e., the limit $lim_{n\rightarrow\infty}\bar{C}(P^n)$
 where $P^n$ is the probability distribution obtained when $n$ runs of the protocol carried out in parallel i.e., when the parties
 receive n inputs and produce $n$ outputs in one go.  A fundamental
 result for this scenario is that $k2^n$ ($k$ a constant) cbits of
 classical communication is required to simulate the correlations
 implied by a n qubit maximally entangled state \cite{brassard}. This
 was followed by a remarkable result due to Toner and Bacon
 \cite{toner} who showed that a single cbit of communication is
 enough (apart from shared random variables) to simulate the
 correlations of a two qubit singlet state. Hence the nonlocality of two spin-$1/2$ particles in the singlet state is one cbit.
 We have later shown  \cite{ahanj1,ahanj2} that for simulating entanglement of
 arbitrary spin S singlet state, communication of
 $n = log_{2}(s +1)$ cbits is enough, provided we confine only to spin measurements.\\Another resource which Alice and Bob
 can use to reproduce the quantum correlations is postselection.
 Here, Alice and Bob are allowed to produce a special outcome of their measurement.
 This corresponds to the physical situation where Alice and Bob's detectors are partially inefficient and sometimes do not click. Gisin and Gisin \cite{gisingisin},
 inspired by Steiner's communication protocol, gave a protocol which simulates quantum correlations
 with shared randomness and with a probability $1/3$ of aborting for either party.\\
 Another fruitful approach to this problem is to use PR-box \cite{scarani}.
 PR-box is a conceptual and mathematical tool developed to study non-locality, first proposed by Popescu and Rohrlich \cite{pr}. It was demonstrated in
 \cite{cerf}, that the correlations of the two qubit singlet can be simulated by supplementing hidden variables ( shared randomness)
 with a single use of the PR-box. Although mathematically based on the Toner and Bacon \cite{toner} result, this work is a major
 conceptual improvement, as the PR-box is a strictly weaker resource than a bit of communication, because it does not allow signaling.\\
 Finally, J. Degorre et.al \cite{j72,j75} have shown that the problem of reproducing the quantum correlation, in the worst case scenario for the two qubit singlet state and in the average scenario for the simulation of traceless binary observables on any bipartite state, with different
 resources, can be reduced to a distributed sampling problem. They have shown that the problem of reproducing the quantum correlations with different resources  (classical communication, PR box, post selection) can be reduced to the problem of Alice and Bob agreeing on a sample from a distribution depending on Alice's input. They introduced a method to carry out this distributed sampling in two steps. The first is the completely local problem of how Alice can sample a biased distribution depending on her input with the help of a shared uniform random source, and the second step is how Alice can share this biased sample with Bob
 by using communication, post-selection or non-local (PR) box (for the two qubit state).\\
  It was soon discovered that for the general case of bipartite qubits in a partially entangled state two
 bits of communication is enough \cite{toner}. Brunner, Gisin and Scarani \cite{brunner05} showed that a single
 use of PR-box is provably not sufficient to simulate some partially entangled two qubit states.
  One may think that the nonlocality of a partially entangled state shouldn't
 be larger than that of maximally entangled state. A reason for this apparently surprising result is that PR-boxes have random marginals,
 ($\langle \alpha \rangle = 0 = \langle \beta \rangle$, a fact which is consistent with two qubit singlet state) while the correlations
 arising from partially entangled quantum states have nontrivial marginals. Thus it appears that it is especially difficult to
 simulate at the same time nonlocal correlations and nontrivial marginals, like these corresponding to partially entangled quantum
 states. Recently, Brunner, Gisin, Popescu and Scarani \cite{bgps}
 (hereafter referred to as BGPS) have given a procedure to simulate
 entanglement in non-maximally entangled states using four PR-boxes and one M-box. The M-box is defined by $a\oplus b =[x\leq y]$ where
 $[x\leq y]$ is the truth value of the predicate $x\leq y$ for given values of $x$ and $y$.
 In order to overcome the above difficulty due to nontrivial marginals, they introduce the concept of correlated local flips, which is independent of whether we use nonlocal boxes or classical
 communication to simulate entanglement. The idea is that some
 nonlocal box, or a classical communication protocol, first simulates
 non-local correlations with trivial marginals and then use local
 flips to bias the marginals.\\
 In this paper, we present a protocol to simulate the quantum  correlations implied by a partially entangled two qubit state, in the worst case scenario,
 by using shared random variables as unit vectors uniformly and independently distributed over the unit sphere in $\mathbb{R}^{4}$ and single use of a M-box and a PR-box. The paper is organized as follows. In Sec II, we recall the BGPS protocol. In Sec III, we present our protocol. Finally we conclude in section IV.

  \section{BGPS protocol}
  We briefly review the BGPS protocol and the idea of the correlated local flips. The problem is to simulate the quantum correlation implied by a general non-maximally entangled two-qubit state
  \begin{equation}\label{psi}
  |\psi\rangle = \cos(\gamma)|00\rangle +\sin(\gamma)|11\rangle ~~~~(0<\gamma<\frac{\pi}{4}).
  \end{equation}
  Alice and Bob perform measurements along directions $\hat{a}$ and $\hat{b}$ on their qubits. Let $\alpha$, $\beta$ denote their outputs respectively.
 For binary outcomes $(\alpha, \beta\in \{-1,+1\})$, the correlations are  conveniently written as
 \begin{equation} \label{joint}
 P(\alpha,\beta | \hat{a},\hat{b})=\frac{1}{4}(1+\alpha
 M_A(\hat{a})+\beta M_B(\hat{b})+\alpha\beta C(\hat{a},\hat{b})) \end{equation} where
 \begin{eqnarray}
 M_A(\hat{a})&=& \sum_{\alpha,\beta}\alpha P(\alpha,\beta |
 \hat{a},\hat{b})\nonumber\\
 M_B(\hat{b})&=& \sum_{\alpha,\beta}\beta P(\alpha,\beta |
 \hat{a},\hat{b})
 \end{eqnarray}
 are the local marginals, and \begin{equation}\label{1}
 C(\hat{a},\hat{b})= \sum_{\alpha,\beta}\alpha \beta P(\alpha,\beta |
 \hat{a},\hat{b})
 \end{equation}
 is the correlation term. Here we shall focus on pure non-maximally entangled states of two qubits.
  Thus the quantum correlation $P_{QM}(\alpha,\beta|\hat{a},\hat{b})$ is given by
 \begin{eqnarray} \label{j7}
 M_A(\hat{a})&=& c a_z \nonumber\\
 M_B(\hat{b})&=&c b_z\nonumber\\
 C(\hat{a},\hat{b})&=& a_z b_z +s(a_xb_x-a_yb_y),
 \end{eqnarray}
 where $c\equiv \cos 2\gamma$ and $s\equiv \sin 2\gamma$.\\
 This gives
 \begin{equation}
 \label{joint1}
 P_{QM}(\alpha,\beta | \hat{a},\hat{b})=\frac{1}{4}(1+\alpha
 c a_z+\beta c b_z+\alpha\beta C(\hat{a},\hat{b}))
 \end{equation}
 where $C(\hat{a},\hat{b})$ is defined in Eq.(\ref{j7}).
  In order to simulate this joint probability, BGPS \cite{bgps} proceed as follows.
 They first set up a procedure ( involving non-local boxes or
 communication) to simulate the joint probability $P_0(\alpha \beta
 |\hat{a},\hat{b}) = \frac{1}{4}[1+ \alpha \beta C_0] $ where $C_0$ is
 the correlation $\langle \alpha \beta \rangle_{0} =\sum \alpha \beta
 P(\alpha, \beta)$  {\it before} flipping.
 Next, they  invoked the local flip operation as follows. Alice
 (Bob) flips the output -1 with probability $f_a(f_b)$ while the output +1 is left untouched. After the local flipping operation with
 probabilities $f_a$ and $f_b$ respectively by Alice and Bob, assuming $f_b \geq f_a $, the joint probability $P_0$ becomes
 \begin{equation}
 P_f(\alpha,\beta | \hat{a},\hat{b})=\frac{1}{4}[1+\alpha f_a+\beta
 f_b+\alpha\beta (f_a + (1-f_b)C_0)].
 \end{equation}
  In order that $P_f$ coincides with $P_{QM}$, they identify $f_a = ca_z$ and $f_b = cb_z$.
 Note that the condition $f_b \geq f_a$ now becomes $b_z \geq a_z$.
 This gives, $C_0 = \hat{a}.\hat{B}$ where $\hat{B} = (sb_x, sb_y,
 b_z - c) / (1 - c b_z)$. If $f_a \geq f_b$, $(a_z \geq b_z)$ , then
 \begin{equation}
 P_f(\alpha,\beta | \hat{a},\hat{b})=\frac{1}{4}[1+\alpha f_a+\beta f_b+\alpha\beta (f_b + (1-f_a)C_0)]. \end{equation}
 where $C_0 = \hat{A}.\hat{b}$ and $\hat{A} = (s a_x, s a_y, a_z- c)
 / (1 - c a_z).$ It is easy to see that $\hat{A}$ and $\hat{B}$ are
 unit vectors. The joint probability $P_0$ to be simulated before
 local flipping operation depends on whether $(b_z \geq a_z)$ or
 $(a_z \geq b_z)$ (via $C_0$). But Alice (Bob) cannot have any information on $b_z$ ($a_z$). In order to pave way through this
 situation BGPS invoke a nonlocal box called M-box which has two real
 inputs $x, y \in [0, 1]$ and binary outputs $m, n \in \{0, 1\}$. The
 M-box is defined by $ m \oplus n = [x \leq y]$ where $[x\leq y]$ is
 the truth value of the predicate $x \leq y $ for given values of $x$
 and $y$.

 \section{The Protocol}
 In this section, we present a protocol for simulating
 entanglement in an arbitrary non maximally entangled two qubit
 quantum state (Eq.(\ref{psi})) in the worst case scenario, which makes a single use of M-box \cite{bgps} and a single use of nonlocal box. The non-local box is used to obtain a random variable whose distribution depends on Alice's input \cite{j72,j75}.\\
 Let Alice and Bob share a  PR-box and a M-box as well as the normalized vectors $\hat{\mu}_1$, $\hat{\mu}_2$, $\hat{\mu}_3$, $\hat{\mu}_4$, $\hat{\mu}_5$, $\hat{\lambda}_0$ and $\hat{\lambda}_1$ independently and uniformly distributed over the unit sphere in $\mathbb{R}^{4}.$
   The protocol runs as follows:\\

  (i) Alice and Bob input $a_z$ and $b_z$ values respectively in
  M-box, which outputs $m$ and $n$ ($m,n\in\{0,1\}$) as above. We replace $m$ and $n$ by $p=2m-1$ and $q=2n-1$ so that $p,q\in\{-1,1\}$. Alice (Bob) gets
  $p(q)$ without knowing   $q(p)$.\\

  (ii) Alice (Bob) prepares the unit vector $\hat{u} \in \mathbb{R}^{4} $ ( $\hat{v} \in \mathbb{R}^{4} $ ) by using  shared random
   vectors $\hat{\mu}_1$, $\hat{\mu}_2$, $\hat{\mu}_3$, $\hat{\mu}_4$, $\hat{\mu}_5$ uniformly and independently distributed over the unit sphere in
   $\mathbb{R}^{4}$ and $p$($q$). These are given by
   \begin{eqnarray}
   \hat{u}&=& \frac{1+p}{2}\hat{u}_1+\frac{1-p}{2}\hat{u}_2\\
   \hat{v}&=& \frac{1+q}{2}\hat{v}_1+\frac{1-q}{2}\hat{v}_2
   \end{eqnarray}
    where
     \begin{eqnarray}
     \label{1}
     \hat{u}_{1,2}& \equiv &(\vec{u}_{1,2},i u_{01,02})\\
     \hat{v}_{1,2}& \equiv &(\vec{v}_{1,2},i v_{01,02}),
      \end{eqnarray}
     where $ i \equiv \sqrt{-1}$ and $u_{0k}$ ($v_{0k}$) is the
   zeroth components of vector ${\hat{u}}_{k}$ (${\hat{v}}_{k}$).
      \begin{eqnarray}
     \label{2}
     \vec{u}_{1,2}&=& sgn(\hat{c}.\hat{\mu}_{1,4})\hat{a}+sgn(\hat{c}.\hat{\mu}_{2,3})\hat{A}\\
     u_{0k}&=& sgn(\hat{c}.\hat{\mu}_{5})|\|\vec{u}_{k}\|^2-1|^{\frac{1}{2}}~~~~~k=1,2 \nonumber\\
     \label{22}
     \vec{v}_{1,2}&=&sgn(\hat{c}.\hat{\mu}_{3,2})\hat{b}+sgn(\hat{c}.\hat{\mu}_{1,4})\hat{B}\\
     v_{0k}&=& |\|\vec{v}_{k}\|^2-1|^{\frac{1}{2}}~~~~~k=1,2.  \nonumber
     \end{eqnarray}
    Here $\rm sgn(x)=1$ for $x\geq 0$ and $\rm sgn(x)=-1$ for $x < 0$ ($ x\in \mathbb{R}$).
     The unit vectors $\hat{A}\in \mathbb{R}^3$ and $\hat{B}\in \mathbb{R}^3$ are defined in the previous section
    and $\hat{c}\in \mathbb{R}^4$ is a fixed unit vector shared between Alice and Bob.\\

    (iii) Now, Alice and Bob use two uniform shared random variables
     $\hat{\lambda}_0\in \mathbb{R}^4$ and $\hat{\lambda}_1\in \mathbb{R}^4$ defined above and make a single use of the non-local (PR) box to output
    \begin{eqnarray}\label{ab}
   \alpha &=& sgn (\hat{u}.\hat{\lambda}_s)\\
   \beta  &=& sgn (\hat{v}.\hat{\lambda}_s).
   \end{eqnarray}
where $\hat{\lambda}_s \in \mathbb{R}^4$ is a random unit vector distributed according to a biased distribution with probability density
\begin{equation}\label{prdf}    
\rho_{\hat{u}}(\hat{\lambda}_{s})=\frac{| \hat{u}.\hat{\lambda}_s|}{2\pi^2}
\end{equation}  
 and is {\it not} a shared random variable between Alice and Bob \cite{j72,j75}.
Note that the distribution $\rho_{\hat{u}}(\hat{\lambda}_{s})$ depends on Alice's input via ${\hat{u}}.$ The corresponding protocol consists of Alice producing $\hat{\lambda}_s$ distributed according to $\rho_{\hat{u}}(\hat{\lambda}_{s})$ (Eq.(\ref{prdf})) and then Alice and Bob make a single use of the non-local (PR) box so as to enable Bob to produce his output as in Eq.(\ref{ab}) without knowing $\hat{\lambda}_s .$ The protocol to do this, making a single use of non-local (PR) box, is given by a trivial modification in the `non-local box' protocol in Ref \cite{j72, 14}.\\

We now find all the relevant averages at this stage of the protocol. Integrating over $\hat{\lambda}_s$ , we obtain:
\begin{eqnarray}
<\alpha \beta>_{\hat{\lambda}_{s}}&=&\int_{\mathbb{S}_3}\rho_{\hat{u}}(\hat{\lambda}_{s})sgn (\hat{u}.\hat{\lambda}_s)sgn (\hat{v}.\hat{\lambda}_s)d\hat{\lambda}_{s}\nonumber\\ &=&\frac{1}{2\pi^2}\int_{\mathbb{S}_3}| \hat{u}.\hat{\lambda}_s|sgn (\hat{u}.\hat{\lambda}_s)
sgn (\hat{v}.\hat{\lambda}_s)d\hat{\lambda}_{s}\nonumber\\&=&\frac{1}{2\pi^2}\int_{\mathbb{S}_3}(\hat{u}.\hat{\lambda}_{s})
sgn (\hat{v}.\hat{\lambda}_s)d\hat{\lambda}_{s}\nonumber\\&=&\frac{1}{2\pi^2}\hat{u}.\int_{\mathbb{S}_3}\hat{\lambda}_s
sgn (\hat{v}.\hat{\lambda}_s)d\hat{\lambda}_{s}\nonumber\\&=&\hat{u}.\hat{v},
\end{eqnarray}
where the the first equality defines $<\alpha \beta>_{\hat{\lambda}_{s}},$ $\mathbb{S}_3$ stands for the unit sphere in $\mathbb{R}^4$ and the final integral is given by $\int_{\mathbb{S}_3}\hat{\lambda}_s sgn (\hat{v}.\hat{\lambda}_s)d\hat{\lambda}_{s}=(2\pi^2)\hat{v}~$  ( see lemma 4 in \cite{j75}).
Also it is easy to check that $<\alpha>_{\hat{\lambda}_s}=0=<\beta>_{\hat{\lambda}_s}.$

The average before the local flipping operation (see below) (denoted $\langle\cdot\rangle_{0}$), is obtained by averaging over $\{\hat{\mu}\}$ ( denoted $\langle\cdot\rangle_{\{\mu\}}$). We get:
$$<\alpha>_0 = <\beta>_0 = 0 $$ and
\begin{eqnarray}
\label{3}
&<&\alpha \beta>_0  = <\hat{u}.\hat{v}>_{\{\mu\}}\nonumber\\&=&\frac{1+p}{2}\frac{1+q}{2}<\hat{u}_1 .\hat{v}_1>_{\{\mu\}}+\frac{1-p}{2}\frac{1-q}{2}<\hat{u}_2 .\hat{v}_2>_{\{\mu\}}\nonumber\\
&=&\frac{1+p}{2}\frac{1-q}{2}<\hat{u}_1 .\hat{v}_2>_{\{\mu\}}+\frac{1-p}{2}\frac{1+q}{2}<\hat{u}_2 .\hat{v}_1>_{\{\mu\}}.\nonumber \\
\end{eqnarray}
From Eq(\ref{1})-Eq(\ref{22}), we can calculate all $ \hat{u}_{1,2} .\hat{v}_{1,2}$. For example:
\begin{eqnarray}
\hat{u}_1 .\hat{v}_1 &=& \vec{u}_1.\vec{v}_1 - u_{01}v_{01}= sgn(\hat{c}.\hat{\mu}_1)sgn(\hat{c}.\hat{\mu}_3)\hat{a}.\hat{b}\nonumber\\& +&
sgn(\hat{c}.\hat{\mu}_2)sgn(\hat{c}.\hat{\mu}_1)\hat{A}.\hat{B}\nonumber\\& +&
sgn(\hat{c}.\hat{\mu}_2)sgn(\hat{c}.\hat{\mu}_3)\hat{A}.\hat{b}\nonumber\\& +&
\hat{a}.\hat{B} \nonumber\\&-& sgn(\hat{c}.\hat{\mu}_5)|\|\vec{u}_{1}\|^2-1|^{\frac{1}{2}}|\|\vec{v}_{1}\|^2-1|^{\frac{1}{2}}.
\end{eqnarray}
  Now by using  $<sgn(\hat{c}.\hat{\mu}_{i})sgn(\hat{c}.\hat{\mu}_{j})>_{\{\mu\}}=\delta_{ij}$ and $<sgn(\hat{c}.\hat{\mu}_{i})>_{\{\mu\}}=0$, we obtain:
\begin{eqnarray}
<\hat{u}_1.\hat{v}_1>_{\{\mu\}}&=&<\hat{u}_2.\hat{v}_2>_{\{\mu\}}=\hat{a}.\hat{B}\\
<\hat{u}_1.\hat{v}_2>_{\{\mu\}}&=&<\hat{u}_2.\hat{v}_1>_{\{\mu\}}=\hat{A}.\hat{b}.
\end{eqnarray}
Substituting these in Eq(\ref{3}), we get:
\begin{equation}
\label{4}
<\alpha \beta>_0  =\frac{1+pq}{2} \hat{a}.\hat{B} +\frac{1-pq}{2} \hat{A}.\hat{b}~.
\end{equation}
(iv) Alice and Bob perform the local flip operation with probabilities $f_a = ca_z$ and $f_b=cb_z$ respectively. We denote the averages after flipping described in section II by $\langle\cdot\rangle_{f}$ and averages in the quantum state $|\psi\rangle$ (Eq.(\ref{psi})) by $\langle\cdot\rangle_{QM}.$ Now, if $f_b>f_a~(b_z>a_z)$ then $p=q$
which implies from Eq.(\ref{4}) that $<\alpha \beta>_0=C_0=\hat{a}.\hat{B}$, so that $<\alpha \beta>_f = ca_z + (1-cb_z)\hat{a}.\hat{B}=<\alpha \beta>_{QM}$.
If $f_a>f_b~(a_z>b_z)$ then $p\neq q$ which implies from Eq.(\ref{4}) that $<\alpha \beta>_0=C_0=\hat{A}.\hat{b}$, so that $<\alpha \beta>_f = cb_z +
 (1-c a_z)\hat{A}.\hat{b}=<\alpha \beta>_{QM}$. The local flipping operation ensures that $<\alpha>_f = f_a =ca_z = <\alpha>_{QM}$ and $<\beta>_f = f_b =cb_z = <\beta>_{QM}$. We see that the protocol simulates the joint probability $P_{QM}(\alpha, \beta|\hat{a},\hat{b}) $ as in Eq.(\ref{joint1}).
 \section{Conclusion}
 We have constructed a protocol to simulate quantum correlation implied by a partially entangled two qubit state, making single use of M-box and nonlocal (PR) box. The non-local box is used to achieve the required (local) outputs by the two parties which depend on a random variable not shared by the parties. This protocol simulates partially entangled two qubit state correlations in the worst case scenario. Our protocol is an improvement on previous protocols which used cbit communication (for example 2 cbit communication in Toner and Bacon model \cite{toner}) and non-local boxes (four PR-box and one M-box in BGPS model \cite{bgps}), as it uses only one M-box and one PR-box which is a weaker resource. Obviously, this is a minimal resource in the worst case scenario.
 \section*{Acknowledgment}
 It is a pleasure to thank N. Gisin, N. Brunner and S. Popescu for very
encouraging and useful correspondence. We thank Prof. R. Simon, G. Kar,
S. Ghosh and P. Rungta for discussions and encouragement. One of
us, PSJ, thanks BCUD (Grant RG-13) for financial support.


\begin{thebibliography}{99}
\bibitem{14} In ref \cite{j72}, the `nonlocal box' protocol makes use of the `choice method' which is proved for 3-D random vectors uniformly and independently distributed over $\mathbb{S}_2$. However, the same protocol and the `choice method' work for 4-D random vectors uniformly and independently distributed on $\mathbb{S}_{3}$ , to give
     $\hat{\lambda}_s \in \mathbb{R}^4$ with the baised distribution as in Eq.(\ref{density}).
\end{thebibliography}
\end{document}